\documentclass[preprint,amsmath,amssymb,11pt,english,aps,showkeys,showpacs]{revtex4}
\usepackage[english]{babel}
\usepackage{graphicx}
\usepackage{graphics}
\usepackage{amsmath}
\usepackage{dcolumn}
\usepackage{amssymb}
\usepackage{bm}
\begin{document}
\title{Effects of a Landau-type quantization induced by the Lorentz symmetry violation on a Dirac field}

\author{R. L. L. Vit\'oria}
\email{ricardo.vitoria@pq.cnpq.br/ricardo-luis91@hotmail.com}
\affiliation{Departamento de F\'isica e Qu\'imica, Universidade Federal do Esp\'irito Santo, Av. Fernando Ferrari, 514, Goiabeiras, 29060-900, Vit\'oria, ES, Brazil.}

\author{H. Belich}
\email{belichjr@gmail.com}
\affiliation{Departamento de F\'isica e Qu\'imica, Universidade Federal do Esp\'irito Santo, Av. Fernando Ferrari, 514, Goiabeiras, 29060-900, Vit\'oria, ES, Brazil.}

\begin{abstract}
Inspired by the extension of the Standard Model, we analyzed the effects of the spacetime anisotropies on a massive Dirac field through a non-minimal CPT-odd coupling in the Dirac equation, where we proposed a possible scenario that characterizes the breaking of the Lorentz symmetry which is governed by a background vector field and induces a Landau-type quantization. Then, in order to generalize our system, we introduce a hard-wall potential and, for a particular case, we determine the energy levels in this background. In addition, at the non-relativistic limit of the system, we investigate the effects of the Lorentz symmetry violation on thermodynamic aspects of the system.
\end{abstract}


\maketitle

\section{Introduction}

Recently, Kosteleck\'y and Samuel \cite{kos} have shown that in the context of string field theory, the Lorentz symmetry (LS) violation governed by tensor fields is natural when the perturbative string vacuum is unstable. Carroll, Field and Jackiw \cite{cfj}, in the context of electrodynamics, have investigated the theoretical and observational consequences of the presence of a background vector field in the modified Cherm-Simons lagrangean, that is, in $(3+1)$-dimensions, which preserves the gauge symmetry, but violates the Lorentz symmetry. One of the purposes of these investigations is the extension of theories and models which may involve the LS violation with the intention of searching for the underlying physical theories that can answer questions that the usual physics can not. In this sense, the Standard Model (SM) has been the target of these extensions characterized by the LS violation which culminated in what we know today as the Extended Standard Model (ESM) \cite{kos1, kos2}. In recent years, the LS violation has been extensively studied in various branches of physics, for example, in magnetic moment generation \cite{bel}, in Rashba spin-orbit interaction \cite{slv}, in Maxwell-Chern-Simons vortices \cite{slv1}, on vortexlike configurations \cite{bel1}, in Casimir effect \cite{slv2, slv3}, in cosmological constraints in electrodynamic \cite{slv4, slv5} and on an analogy of the quantum hall conductivity \cite{slv6}.

In recent years, the LS violation has been applied in quantum mechanics systems. In the non-relativistic limit, for example, there are studies in an Aharonov-Bohm-Casher system \cite{bel2}, in quantum holonomies \cite{bb}, in a Dirac neutral particle inside a two-dimensional quantum ring \cite{bb2}, on a spin-orbit coupling for a neutral particle \cite{bb3} and in a system under the influence of a Rashba-type coupling induced \cite{bb4}. In the relativistic {\bf{case}} there are studies on EPR correlations \cite{bb5}, in geometric quantum phases \cite{bb6}, in the Landau-He-McKellar-Wilkens quantization and bound states solutions for a Coulomb-like potential \cite{bb7}, in a quantum scattering \cite{bb8} and on a scalar field \cite{bb9, bb10, me, me1, me2, me2a, me2b}. One point that has not been dealt with in the literature is the relativistic Landau-type quantization induced by the LS violation under a Dirac field.

Recently, thermodynamic properties of quantum systems have been investigated and, consequently, can be found in the literature. Here, we can cite some examples of these studies, for example, in diatomic molecule systems \cite{pt}, in a neutral particle system in the presence of topological defects in magnetic cosmic string background \cite{pt1}, in exponential-type molecule potentials \cite{pt2}, on the a 2D charged particle confined by a magnetic and Aharonov-Bohm flux fields under the radial scalar power potential \cite{pt3}, on the Dirac and Klein-Gordon oscillators in the anti-de Sitter space \cite{pt4}, on the Klein-Gordon oscillator in the frame work of generalized uncertainty principle \cite{pt5} and the on the harmonic oscillator in an environment with a pointlike defect \cite{pt6}.

In this paper, we investigate the effects of the LS violation on a Dirac field, where the spacetime anisotropies are governed by the presence of a vector background field inserted in the Dirac equation via non-minimal coupling that, for a possible scenery in this background characterized by a particular electromagnetic field configuration, it is possible to induce a relativistic analogue of the Landau quantization. Further, we analyse the Dirac field under the influence of a hard-wall potential in this LS violation background for a particular case. In addition, inspired by Refs. \cite{pt, pt1, pt2, pt3, pt4, pt5, pt6}, we investigate the thermodynamic properties of our more general system investigated at low energies and analyze the LS violation effects on some thermodynamic magnitudes.

The structure of this paper is as follows: in the Sec. II we introduce a LS violation background defined by a vector field that governs the spacetime anisotropies, and thus, establish a possible scenario of the LS violation that induces a relativistic analogue of the Landau quantization; in the Sec. III we inserted into the system a hard-wall potential and, for a particular case, we determined the relativistic energy levels in this LS violation background; in the Sec. IV, through the Landau-type non-relativistic energy levels, we investigated some thermodynamic aspects under the effects of the LS violation; in the Sec. V we present our conclusions.

\section{Landau-type quantization induced by the Lorentz symmetry violation}

In this section, we investigated the influence of the spacetime anisotropies under a Dirac field, where we obtain the relativistic Landau-type energy levels induced by the LS violation effects. Recently, in low energy scenarios, the Landau-type quantization induced by the LS violation has been investigated \cite{bb11, book}. These analyzes are possible through couplings non-minimal in wave equations \cite{bb9, book, bel12, bb13}, where this procedure carries the information that there are privileged directions in the spacetime to which they characterize the LS violation. From the mathematical point of view, this information is given through the presence of constant background fields of vector or tensor nature \cite{bel13}. Then, based on Refs. \cite{book, bb13}, consider the non-minimum coupling defined by (with $c=\hbar=k_{B}=1$) $i\gamma^{\mu}\partial_{\mu}\rightarrow i\gamma^{\mu}\partial_{\mu}-g\gamma^{\mu}\tilde{F}_{\mu\nu}v^{\nu}$, where $g$ is coupling constant, $\gamma^{\mu}$ are Dirac matrices
\begin{eqnarray}\label{eq01}
\gamma^0=\left[
  \begin{array}{cc}
    I & 0 \\
    0 & -I \\
  \end{array}
\right]; \ \ \
\gamma^{i}=\left[
  \begin{array}{cc}
    0 & \sigma^{i} \\
    -\sigma^{i} & 0 \\
  \end{array}
\right]; \ \ \
\gamma^{i}=\beta\alpha^{i}=\gamma^{0}\alpha^{i},
\end{eqnarray}
with $\gamma^{\mu}\gamma^{\nu}+\gamma^{\nu}\gamma^{\mu}=2\eta^{\mu\nu}$ and $\sigma^{i}$ are the Pauli matrices, $\tilde{F}_{\mu\nu}(x)=\frac{1}{2}\epsilon_{\mu\nu\alpha\beta}F^{\alpha\beta}(x)$ is dual electromagnetic tensor, where $F_{0i}=-F_{i0}=-E_{i}$ and $F_{ij}=-F_{ji}=\epsilon_{ijk}B^{k}$, and $v^{\mu}$ is background vector field that governs the LS violation. This nonminimal coupling is inspired by the gauge sector of the SME \cite{kos1, kos2}, which is known as the non-minimal CPT-odd coupling, as it, in addition to breaking the Lorentz symmetry, also breaks the CPT symmetry \cite{book, rl}. In the way, the Dirac equation in any orthogonal system with the non-minimal coupling is \cite{book}
\begin{eqnarray}\label{eq02}
i\gamma^{\mu}D_{\mu}\Psi+\frac{1}{2}\sum_{j=0}^{3}\gamma^{j}\left[D_{j}\ln\left(\frac{h_1h_2h_3}{h_j}\right)\right]\Psi
-g\gamma^{\mu}\tilde{F}_{\mu\nu}v^{\nu}\Psi-m\Psi=0,
\end{eqnarray}
where $D_{\mu}=\frac{1}{h_{(\mu})}\partial_{\mu}$ is the derivative of the corresponding coordinate system and $h_{(\mu)}$ is the parameter which corresponds to the scale factors of this coordinate system. This means that a new contribution to the Dirac equation can stem from the coordinate system in an analogous way to the well-known coupling between spinors and curvature discussed in quantum field theory in curved space \cite{birrell}. In this paper, we are working with the Minkowski spacetime in cylindrical coordinates
\begin{eqnarray}\label{eq03}
ds^2=-dt^2+d\rho^2+\rho^2d\varphi^2+dz^2,
\end{eqnarray}
then, the scale factors are $h_0=h_1=h_3=1$ and $h_2=\rho$. In this way, the Eq. (\ref{eq02}) is rewritten as follows
\begin{eqnarray}\label{eq04}
i\gamma^{0}\frac{\partial\Psi}{\partial t}&+&i\gamma^{1}\left(\frac{\partial}{\partial\rho}+\frac{1}{2\rho}\right)\Psi
+i\frac{\gamma^{2}}{\rho}\frac{\partial\Psi}{\partial\varphi}+i\gamma^{3}\frac{\partial\Psi}{\partial z}+gv^0\gamma^0\vec{\alpha}.\vec{B}\Psi-g\gamma^{0}\vec{v}.\vec{B}\Psi \nonumber \\
&-&g\gamma^0\vec{\alpha}.(\vec{v}\times\vec{E})\Psi-m\Psi=0.
\end{eqnarray}

Now, let us consider a LS violation possible scenario where we have the background field and the presence of an electric field given by
\begin{eqnarray}\label{eq05}
v^{\mu}=(0,\vec{v}=a\hat{z}); \ \ \ \vec{E}=\frac{\vartheta}{2}\rho\hat{\rho},
\end{eqnarray}
where $a=$const. and $\vartheta$ is a parameter associated to the uniform volumetric distribution of electric charges. We can note that the possible scenario of the LS violation background field configuration given in Eq. (\ref{eq05}) induces an analogue to the quantization of Landau \cite{landau}, that is, we have a vector potential $\vec{\mathcal{A}}=\vec{v}\times\vec{E}$ which it gives a uniform ``magnetic field'' $\vec{\mathcal{B}}=\nabla\times\vec{\mathcal{A}}=\mathcal{B}_0\hat{z}=a\vartheta\hat{z}$. In addition, the electric field configuration given in Eq. (\ref{eq05}) has been studied in Landau levels induced by magnetic dipole moment \cite{bf}, in Landau levels induced by electric dipole moment \cite{mde, mde1, bf1} and in LS violation possible scenarios \cite{bb4, bb10, me1}. In this way, the Dirac equation (\ref{eq04}) becomes
\begin{eqnarray}\label{eq06}
i\gamma^{0}\frac{\partial\Psi}{\partial t}+i\gamma^{1}\left(\frac{\partial}{\partial\rho}+\frac{1}{2\rho}\right)\Psi
+i\frac{\gamma^{2}}{\rho}\frac{\partial\Psi}{\partial\varphi}+i\gamma^{3}\frac{\partial\Psi}{\partial z}-\frac{ga\vartheta}{2}\rho\gamma^2\Psi-m\Psi=0.
\end{eqnarray}

Assuming stationary state solutions and that the linear momentum $\hat{p}_z=-i\frac{\partial}{\partial z}$ and angular momentum $\hat{J}_z=-i\frac{\partial}{\partial\varphi}$ operators commute with the hamiltonian operator, we have the following solution
\begin{eqnarray}\label{eq07}
\Psi(\rho,\varphi,z,t)=e^{-i\mathcal{E}t}e^{i\left(l+\frac{1}{2}\right)\varphi}e^{ikz}\left[
                                                                                        \begin{array}{c}
                                                                                          \psi_{+}(\rho) \\
                                                                                          \psi_{-}(\rho) \\
                                                                                        \end{array}
                                                                                      \right],
\end{eqnarray}
where $l=0,\pm1,\pm2,\pm3,\ldots$ are the eigenvalues of angular momentum and $-\infty<k<\infty$. In this way, by substituting the Eq. (\ref{eq07}) into the Eq. (\ref{eq06}) and considering the matrices given in Eq. (\ref{eq01}), we obtain the two coupled equations below
\begin{eqnarray}\label{eq08}
(\mathcal{E}-m)\psi_{+}&+&i\sigma^{1}\left(\frac{d}{d\rho}+\frac{1}{2\rho}\right)\psi_{-}-\frac{1}{\rho}\left(l+\frac{1}{2}\right)\sigma^2\psi_{-}
-k\sigma^{3}\psi_{-}-\frac{ag\vartheta}{2}\sigma^{2}\psi_{-}=0; \nonumber \\
(\mathcal{E}+m)\psi_{-}&+&i\sigma^{1}\left(\frac{d}{d\rho}+\frac{1}{2\rho}\right)\psi_{+}-\frac{1}{\rho}\left(l+\frac{1}{2}\right)\sigma^2\psi_{+}
-k\sigma^{3}\psi_{+}-\frac{ag\vartheta}{2}\sigma^{2}\psi_{+}=0.
\end{eqnarray}

By multiplying first equation by $(\mathcal{E}+m)$ and using the second equation we obtain
\begin{eqnarray}\label{eq09}
\frac{d^2\psi_{+}}{d\rho^2}&+&\frac{1}{\rho}\frac{d\psi_{+}}{d\rho}-\left[\left(l+\frac{1}{2}\right)^2+\frac{1}{4}
-\left(l+\frac{1}{2}\right)\sigma^{3}\right]\frac{\psi_{+}}{\rho^2}-\frac{(ag\vartheta)^2}{4}\rho^2\psi_{+} \nonumber \\
&+&\left[\mathcal{E}^2-m^2-k^2-ag\vartheta\left(l+\frac{1}{2}+\frac{\sigma^{3}}{2}\right)\right]\psi_{+}=0.
\end{eqnarray}
An analog equation can be found for $\psi_{-}$. We can simplify the differential equations for $\psi_{+}$ and $\psi_{-}$, since the $\psi$ is an eigenfunction of $\sigma^{3}$, whose eigenvalues are $s=\pm1$. Thereby, we write $\sigma^{3}\psi_s=\pm\psi_s=s\psi_s$. Hence, we can represent the two differential equations in a single radial differential equation:
\begin{eqnarray}\label{eq10}
\frac{d^2\psi_{s}}{d\rho^2}&+&\frac{1}{\rho}\frac{d\psi_{s}}{d\rho}-\left[l+\frac{1}{2}(1-s)\right]^2\frac{\psi_{s}}{\rho^2}
-\frac{(ag\vartheta)^2}{4}\rho^2\psi_{s} \nonumber \\
&+&\left[\mathcal{E}^2-m^2-k^2-ag\vartheta\left(l+\frac{1}{2}(1+s)\right)\right]\psi_{s}=0.
\end{eqnarray}

We proceed with a change of variables given by $\varrho=\frac{ag\vartheta}{2}\rho^2$, and thus, we rewrite the Eq. (\ref{eq10}) in the form
\begin{eqnarray}\label{eq11}
\frac{d^2\psi_{s}}{d\varrho^2}+\frac{1}{\varrho}\frac{d\psi_{s}}{d\varrho}-\frac{l_{s}^2}{4\varrho^2}\psi_{s}+\frac{\alpha^2_{s}}{\varrho}\psi_{s}
-\frac{1}{4}\psi_{s}=0,
\end{eqnarray}
where we define the parameters
\begin{eqnarray}\label{eq12}
\alpha^2_{s}=\frac{1}{2ag\vartheta}\left[\mathcal{E}^2-m^2-k^2-ag\vartheta\left(l+\frac{1}{2}(1+s)\right)\right]; \ \ \ l_{s}^2=\left[l+\frac{1}{2}(1-s)\right]^2.
\end{eqnarray}

By imposing that $\psi_{s}\rightarrow0$ when $\varrho\rightarrow0$ and $\varrho\rightarrow\infty$, we can write $\psi_{s}$ in terms of an unknown function $F(\varrho)$ as follows:
\begin{eqnarray}\label{eq13}
\psi_{s}=\varrho^{\frac{|l_{s}|}{2}}e^{-\frac{1}{2}\varrho}F(\varrho).
\end{eqnarray}

Then, by substituting the Eq. (\ref{eq13}) into the Eq. (\ref{eq11}), we have the second order differential equation
\begin{eqnarray}\label{eq14}
\varrho\frac{d^2F}{d\varrho^2}+(|l_{s}|+1-\varrho)\frac{dF}{d\varrho}-\left(\frac{|l_{s}|}{2}+\frac{1}{2}-\alpha^2_{s}\right)F=0.
\end{eqnarray}

The Eq. (\ref{eq14}) is called in the literature as the confluent hypergeometric equation and the function $F(\varrho)=_{1}F_1(a,b;\varrho)$, with $a=\frac{|l_{s}|}{2}+\frac{1}{2}-\alpha^2_{s}$ and $b=|l_{s}|+1$, is the confluent hypergeometric function \cite{arf}. It is well known that the confluent hypergeometric series becomes a polynomial of degree $n$ by imposing that $a=-n$, where $n=0,1,2,3,\ldots$. With this condition, we obtain
\begin{eqnarray}\label{eq15}
\mathcal{E}_{k,l,n}=\pm\sqrt{m^2+k^2+2m\omega\left[n+\frac{1}{2}\left|l+\frac{1}{2}(1-s)\right|+\frac{1}{2}\left(l+\frac{1}{2}(1-s)\right)+\frac{(1+s)}{2}\right]},
\end{eqnarray}
where $\omega=\frac{ag\vartheta}{m}$ is the cyclotron frequency. The Eq. (\ref{eq15}) represents the energy levels of the system of a Dirac field in an anisotropic spacetime, where the anisotropies are governed by a background vector field present in a field configuration which provides a relativistic analogue to Landau quantization. We can note that the effects of the LS violation influence the energy profile of the system through the cyclotron frequency which is defined in terms of the parameters associated with the Lorentz symmetry violation, $g$, $a$ and $\vartheta$. We can note that by taking $g\rightarrow0$ we recover energy of a free Dirac field in the Minkowski spacetime.

\subsection{Nonrelativistic limit}

To analyze the nonrelativistic limit of the energy levels, we can write the Eq. (\ref{eq15}) in the form
\begin{eqnarray}\label{eq16}
\mathcal{E}_{k,l,n}=m\left[1+\frac{k^2}{m^2}+\frac{2\omega}{m}\left(n+\iota+\frac{1+s}{2}\right)\right]^{\frac{1}{2}},
\end{eqnarray}
where $\iota=\frac{1}{2}(|l_{s}|+l_s)$. Then, by using the approximation $(1+x)^{\kappa}\simeq1+\kappa x$ in the Eq. (\ref{eq16}) and making $\varepsilon_{k,l,n}=\mathcal{E}_{k,l,n}-m$, where $\varepsilon_{k,l,n}\ll m$, we obtain the expression
\begin{eqnarray}\label{eq17}
\varepsilon_{k,l,n}=\frac{k^2}{2m}+\omega\left[n+\frac{1}{2}\left|l+\frac{1}{2}(1-s)\right|+\frac{1}{2}\left(l+\frac{1}{2}(1-s)\right)+\frac{1+s}{2}\right],
\end{eqnarray}
which represents the Landau-type energy levels of a Dirac particle at low energies in a LS violation possible scenario. We can see that the energy profile of the system is influenced by the LS violation through the cyclotron frequency which is defined in terms of the parameters associated with the LS violation background, $a$, $g$ and $\vartheta$. For the particular case where $k=0$ we recover the result obtained in the Ref. \cite{book}. In addition, by making $g\rightarrow0$ implies in $\omega\rightarrow0$, where we obtain free energy of a Dirac particle in an isotropic environment.

\section{Effects of a hard-wall potential}

In this section, we confine the system described in the previous section to a region where the hard-wall potential is present. This confining potential is characterized by the following boundary condition:
\begin{eqnarray}\label{eq18}
\psi_{s}(\rho_0)=0,
\end{eqnarray}
which means that the radial wave function vanishes at a fixed radius $\rho_0$. The hard-wall potential has been studied in noninertial effects \cite{pr, pr1, pr2, pr3}, in relativistic scalar particle systems in a spacetime with a spacelike dislocation \cite{me3}, in a geometric approach to confining a Dirac neutral particle in analogous way to a quantum dot \cite{pr4}, in a Landau-Aharonov-Casher system \cite{pr5} and on a harmonic oscillator in an elastic medium with a spiral dislocation \cite{pr6}. Then, let us consider the particular case which $\alpha^2_{s}\gg1$, while the $\rho_0$ and the parameter of the confluent hypergeometric function $b$ are fixed and that the parameter of the confluent hypergeometric function $a$ to be large. In the way, a confluent hypergeometric function can be written in the form \cite{me3, abr, kb}:
\begin{eqnarray}\label{eq19}
\ _{1}F_1(a,b,\varrho_0)\propto\cos\left(\sqrt{2\varrho_0(b-2a)}+\frac{\pi}{4}-\frac{b\pi}{2}\right).
\end{eqnarray}

By substituting the Eqs. (\ref{eq13}) and (\ref{eq19}) into Eq. (\ref{eq18}), we obtain
\begin{eqnarray}\label{eq20}
\mathcal{E}_{k,l,n}\approx\pm\sqrt{m^2+k^2+m\omega\left[l+\frac{1}{2}(1-s)+s\right]
+\frac{\pi^2}{\rho_0^2}\left[n+\frac{1}{2}\left|l+\frac{1}{2}(1-s)\right|+\frac{3}{4}\right]^2},
\end{eqnarray}
with $n=0,1,2,\ldots$. The Eq. (\ref{eq20}) represents the energy levels of a Dirac field subject to the effects of a Landau-type quantization induced by the LS violation effects and under the effects of a hard-wall potential. By comparing the Eqs. (\ref{eq15}) and (\ref{eq20}), we can note that the presence of the hard-wall potential in the system modifies the relativistic energy levels. This modification can be seen through the degenrescence that is broken and the main quantum number that now has a quadratic character in contrast to the previous section. In addition, by taking $g\rightarrow0$ we obtain the energy spectrum of a Dirac field subject to a hard-wall confining potential in the Minkowski spacetime.

\subsection{Nonrelativistic limit}

In this subsection we are interested in obtaining the nonrelativistic energy levels of a Dirac field subject to the Landau-type quantization induced by the LS violation effects and the hard-wall potential given in Eq. (\ref{eq18}). In this sense, we can write Eq. (\ref{eq20}) as follows
\begin{eqnarray}\label{eq21}
\mathcal{E}_{k,l,n}\approx m\sqrt{1+\frac{k^2}{m^2}+\frac{\omega}{m}\left[l+\frac{1}{2}(1-s)+s\right]
+\frac{\pi^2}{m^2\rho_{0}^2}\left[n+\frac{1}{2}\left|l+\frac{1}{2}(1-s)\right|+\frac{3}{4}\right]^2}.
\end{eqnarray}
Then, by following the same steps from the Eq. (\ref{eq16}) to the Eq. (\ref{eq17}), we have
\begin{eqnarray}\label{eq22}
\varepsilon_{k,l,n}\approx \frac{k^2}{2m}+\frac{\omega}{2}\left[l+\frac{1}{2}(1-s)+s\right]+\frac{\pi^2}{2m\rho_0^2}\left[n+\frac{1}{2}\left|l+\frac{1}{2}(1-s)\right|+\frac{3}{4}\right]^2.
\end{eqnarray}

The Eq. (\ref{eq22}) represents the non-relativistic energy levels of a Dirac particle under the effects of a Landau-type quantization and confined in a region where a hard-wall potential is present. Note that the energy profile of the system at low energies is influenced by the LS violation and the hard-wall potential by the presence of the parameters $a$, $g$, $\vartheta$ and $\rho_0$, respectively. In addition, in contrast to the Eq. (\ref{eq17}), the main quantum number of the energy spectrum given in Eq. (\ref{eq22}) has quadratic character. In addition, the degeneracy is broken. These latter two effects are caused by the presence of the hard-wall potential. We can observe that by making $g\rightarrow0$ into Eq. (\ref{eq22}) we obtain the energy spectrum of a Dirac particle under confining effects of a hard-wall potential.

\section{Thermodynamic properties}

In order to obtain the thermodynamic magnitudes of a Dirac particle in a LS violation background, we converged, first, to define the partition function from the Landau-type energy levels given in Eq. (\ref{eq17}) considering the quantum numbers $l$, $k$ and $s$ fixed. Then, from now on, let us consider $N$ Dirac particles in a LS violation background with energy given in Eq. (\ref{eq17}), localized and noninteracting, but now in contact with a thermal reservoir at the finite temperature T. The microscopic states of this system are characterized by the set of principal quantum numbers $\{n_{1},n_{2},n_{3},\ldots,n_{N}\}$ \cite{sal}. Therefore, given a microscopic state $n_j$, the energy of that state can be written in the form
\begin{eqnarray}\label{eq23}
\varepsilon_{n_{j}}\{n_j\}=\sum_{j=1}^{N}\left(n_{j}+\frac{\delta_s}{2}\right)\omega,
\end{eqnarray}
where
\begin{eqnarray}\label{eq24}
\delta_s=\frac{k^2}{m\omega}+\left|l+\frac{1}{2}(1-s)\right|+\left(l+\frac{1}{2}(1-s)\right)+1+s.
\end{eqnarray}
Therefore, the canonical partition function is given by \cite{pt1, sal}
\begin{eqnarray}\label{eq25}
Z=\sum_{n_j}e^{-\beta\varepsilon_{n_j}}=\sum_{n_1,n_2,n_3,\ldots,n_N}e^{-\sum_{j=1}^{N}\left(n_{j}+\frac{\delta_{s}}{2}\right)\beta\omega},
\end{eqnarray}
where $\beta=T^{-1}$. Since there are no terms of interaction, the multiple sum is factorized, source damage to a very simple expression, that is,
\begin{eqnarray}\label{eq26}
Z=\left[\sum_{n=0}^{\infty}e^{-\left(n+\frac{\delta_s}{2}\right)\beta\omega}\right]^{N}=Z_{1}^{N},
\end{eqnarray}
where
\begin{eqnarray}\label{eq27}
Z_{1}=\sum_{n=0}^{\infty}e^{-\left(n+\frac{\delta_s}{2}\right)\beta\omega}=\frac{e^{-\frac{\delta_s\beta\omega}{2}}}{1-e^{-\beta\omega}},
\end{eqnarray}
is the partition function of a single Dirac particle with energy levels given in Eq. (\ref{eq17}).

With the partition function (\ref{eq27}) in hand, we can determine thermodynamic quantities. We start with the free energy of Helmhotz by particle, which is given by the expression \cite{sal}
\begin{eqnarray}\label{eq28}
f&=&-\frac{1}{\beta}\displaystyle \lim_{N \to \infty}\frac{1}{N}\ln Z=-\frac{1}{\beta}\ln Z_1; \nonumber \\
f&=&\frac{k^2}{2m}+\omega\left[\frac{1}{2}\left|l+\frac{1}{2}(1-s)\right|+\frac{1}{2}\left(l+\frac{1}{2}(1-s)\right)+\frac{1+s}{2}\right]
+T\ln(1-e^{-\frac{\omega}{T}}).
\end{eqnarray}

From Eq. (\ref{eq28}), we can calculate all the thermodynamic properties of the system. In particular, entropy by particle $s(T)$ as a function of temperature
\begin{eqnarray}\label{eq29}
s=-\frac{df}{dT}=-\ln(1-e^{-\frac{\omega}{T}})+\frac{\omega}{T}\left(\frac{e^{-\frac{\omega}{T}}}{1-e^{-\frac{\omega}{T}}}\right),
\end{eqnarray}
which we can obtain the specific heat
\begin{eqnarray}\label{eq30}
c=T\frac{ds}{dT}=\left(\frac{\omega}{T}\right)^2\frac{e^{-\frac{\omega}{T}}}{(1-e^{-\frac{\omega}{T}})^2}.
\end{eqnarray}

In addition, we can also calculate the intern energy by particle $u(T)$ as a function of temperature
\begin{eqnarray}\label{eq31}
u&=&-\frac{1}{N}\frac{d}{d\beta}\ln Z=-\frac{d}{d\beta}\ln Z_1; \nonumber \\
u&=&\frac{k^2}{2m}+\omega\left[\frac{1}{2}\left|l+\frac{1}{2}(1-s)\right|+\left(l+\frac{1}{2}(1-s)\right)+\frac{1+s}{2}\right]
+\frac{\omega}{(e^{\frac{\omega}{T}}-1)}.
\end{eqnarray}

We note that the thermodynamic quantities calculated and defined in Eqs. (\ref{eq28}), (\ref{eq29}), (\ref{eq30}) and (\ref{eq31}) are influenced by the LS violation. This influence is due to the presence of the cyclotron frequency in the definition of the partition function (\ref{eq27}) which, in turn, gives us the free energy of Helmhotz by particle (\ref{eq28}) from which we can describe the thermodynamic properties of the system, since the cyclotron frequency $\omega$ is defined in terms of parameters associated with the LS violation in a possible low energy scenario, $a$, $g$ and $\vartheta$.

\section{Conclusion}

We have analyzed a Dirac field in an anisotropic spacetime where there is the presence of a background vector field that governs the LS violation. Then, for our analysis, we have chosen a particular case which characterizes a possible scenario of LS violation that induces a relativistic Landau-type quantization, and then, with this information coming from this background, we calculate the relativistic Landau-type energy levels of a Dirac field in this possible scenario. In addition, we confine the Dirac field in a region where a hard-wall potential is present and, for a particular case, we define the relativistic energy profile of the system, which is modified due to the presence of the hard-wall potential and, consequently, the degeneracy is broken. In both cases, we calculated the energy levels in the low energy limit, where it is possible to notice that there are influences of the LS violation on the energy profiles of the analyzed systems.

With the Landau-type energy levels in a possible scenario of LS violation at low energies, through the partition function and, consequently, of the free energy of Helmholtz by particle, we can calculate thermodynamic quantities of this system, which are influenced by the LS violation background. This influence is due to the dependence of the thermodynamic quantities of the cyclotron frequency, which in turn is defined in terms of parameters associated with the LS violation.

\acknowledgments{The authors would like to thank CNPq (Conselho Nacional de Desenvolvimento Cient\'ifico e Tecnol\'ogico - Brazil) for financial support. R. L. L. Vit\'oria was supported by the CNPq project No. 150538/2018-9.}


\begin{thebibliography}{99}

\bibitem{kos} V. A. Kosteleck\'y, S. Samuel, Phys. Rev. D {\bf{39}}, 683 (1989)

\bibitem{cfj} S. M. Carroll, G. B. Field, R. Jackiw, Phys. Rev. D {\bf{41}}, 1231 (1990)

\bibitem{kos1} D. Colladay, V. A. Kosteleck\'y, Phys. Rev. D {\bf{55}}, 6760 (1997)

\bibitem{kos2} D. Colladay, V. A. Kosteleck\'y, Phys. Rev. D {\bf{58}}, 116002 (1998)

\bibitem{bel} H. Belich, L. P. Collato, T. Costa-Soares, J. A. Helay\"el-Neto, M. T. D. Orlando, Eur. Phys. J. C {\bf{62}}, 425 (2009)

\bibitem{slv} M. A. Ajaib, Int. J. Mod. Phys. A {\bf{27}}, 1250139 (2012)

\bibitem{slv1} R. Casana, M. M. Ferreira Jr., E. da Hora, A. B. F. Neves, Eur. Phys. J. C {\bf{74}}, 3064 (2014)

\bibitem{bel1} H. Belich, F. J. L. Leal, H. L. C. Louzada, M. T. D. Orlando, Phys. Rev. D {\bf{86}}, 125037 (2012)

\bibitem{slv2} M. B. Cruz, E. R. Bezerra de Mello, A. Y. Petrov, Phys. Rev. D {\bf{96}}, 045019 (2017)

\bibitem{slv3} M. B. Cruz, E. R. Bezerra de Mello, A. Y. Petrov, Mod. Phys. Lett. A {\bf{33}}, 1850115 (2018)

\bibitem{slv4} V. A. Kosteleck\'y, M. Mewes, Phys. Rev. Lett. {\bf{87}}, 251304 (2001)

\bibitem{slv5} V. A. Kosteleck\'y, M. Mewes, Phys. Rev. D {\bf{66}}, 056005 (2002)

\bibitem{slv6} L. R. Ribeiro, E. Passos, C. Furtado, J. Phys. G: Nucl. Part. Phys. {\bf{39}}, 105004 (2012)

\bibitem{bel2} H. Belich, E. O. Silva, M. M. Ferreira Jr., M. T. D. Orlando, Phys. Rev. D {\bf{83}}, 125025 (2011)

\bibitem{bb} K. Bakke, H. Belich, J. Phys. G: Nucl. Part. Phys. {\bf{40}}, 065002 (2013)

%\bibitem{bb1} K. Bakke, H. Belich, J. Phys. G: Nucl. Part. Phys. {\bf{39}}, 085001 (2012)

\bibitem{bb2} K. Bakke, H. Belich, Eur. Phys. J. Plus {\bf{129}}, 147, (2014)

\bibitem{bb3} H. Belich, K. Bakke, Int. J. Mod. Phys. A {\bf{30}}, 1550136, (2015)

\bibitem{bb4} K. Bakke, H. Belich, Ann. Phys. {\bf{354}}, 1 (2015)

\bibitem{bb5} H. Belich, K. Bakke, C. Furtado, Eur. Phys. J. C {\bf{75}}, 410 (2015)

\bibitem{bb6} K. Bakke, H. Belich, Int. J. Mod. Phys. A {\bf{30}}, 1550197 (2015)

\bibitem{bb7} K. Bakke, H. Belich,  Ann. Phys. {\bf{333}}, 272 (2013)

\bibitem{bb8} H. F. Mota, H. Belich, K. Bakke, Int. J. Mod. Phys. A {\bf{32}}, 1750140 (2017)

\bibitem{bb9} K. Bakke, H. Belich, Ann. Phys. {\bf{360}}, 596 (2015)

\bibitem{bb10} K. Bakke, H. Belich, Ann. Phys. {\bf{373}}, 115 (2016)

\bibitem{me} R. L. L. Vit\'oria, H. Belich, K. Bakke, Eur. Phys. J. Plus {\bf{132}}, 25 (2017)

\bibitem{me1} R. L. L. Vit\'oria, H. Belich, K. Bakke, Adv. High Ener. Phys. {\bf{2017}}, 6893084 (2017)

\bibitem{me2} R. L. L. Vit\'oria, K. Bakke, H. Belich, Ann. Phys. {\bf{399}}, 117 (2018)

\bibitem{me2a} R. L. L. Vit\'oria, H. Belich, Eur. Phys. J. C {\bf{78}}, 999 (2018)

\bibitem{me2b} R. L. L. Vit\'oria, H. Belich, Adv. High Ener. Phys. {\bf{2019}}, 1248393 (2019)

\bibitem{pt} X.-Qin Song, C.-Wen Wang, C.-Sheng Jia, Chem. Phys. Lett. {\bf{673}}, 50 (2017)

\bibitem{pt1} H. Hassanabadi, M. Hosseinpour, Eur. Phys. J. C {\bf{76}}, 553 (2016)

\bibitem{pt2} A. N. Ikot, B. C. Lutfuoglu, M. I. Ngwueke, M. E. Udoh, S. Zare, H. Hassanabadi, Eur. Phys. J. Plus {\bf{131}}, 419 (2016)

\bibitem{pt3} M. Eshghi, H. Mehraban, Eur. Phys. J. Plus {\bf{132}}, 121 (2017)

\bibitem{pt4} B. Hamil, M. Merad, Eur. Phys. J. Plus {\bf{133}}, 174 (2018)

\bibitem{pt5} B. Khosropour, Indian J. Phys. {\bf{92(1)}}, 43 (2018)

\bibitem{pt6} R. L. L. Vit\'oria, H. Belich, Phys. Scr. {\bf{94}}, 125301 (2019)

\bibitem{bb11} K. Bakke, H. Belich, J. Phys. G: Nucl. Part. Phys. {\bf{42}}, 095001 (2015)

\bibitem{book} K. Bakke, H. Belich, {\it{Spontaneous Lorentz Symmetry Violation and Low Energy Scenarios}} (LAMBERT Academic Publishing, Saarbr\"ucken, 2015)

\bibitem{bel12} H. Belich, T. Costa-Soares, M. M. Ferreira Jr., J. A. Helay\"el-Neto, Eur. Phys. J. C {\bf{41}}, {\bf{421}} (2005)

\bibitem{bb13} K. Bakke, H. Belich, Eur. Phys. J. Plus {\bf{127}}, 102 (2012)

\bibitem{bel13} H. Belich, T. Costa-Soares, M. A. Santos, M. T. D. Orlando, Rev. Bras. Ensino Fís. {\bf{29}}, 1 (2007)

\bibitem{rl} R. Lehnert, Hyperfine Interact {\bf{193}}, 275 (2009)

\bibitem{birrell} N. D. Birrell, P. C. W. Davies, {\it{Quantum Fields in Curved Space}}, (Cambridge University Press, Cambridge, UK, 1982)

\bibitem{landau} L. D. Landau, E. M. Lifshitz, {\it{Quantum Mechanics: The Nonrelativistic Theory}}, 3rd edn. (Pergamon, Oxford, 1977)

\bibitem{bf} K. Bakke, C. Furtado, Phys. Rev. D {\bf{82}}, 084025 (2010)

\bibitem{mde} C. Furtado, J. R. Nascimento, L. R. Ribeiro, Phys. Lett. A {\bf{358}}, 336 (2006)

\bibitem{mde1} L. R. Ribeiro, Claudio Furtado, J. R. Nascimento, Phys. Lett. A {\bf{348}}, 135 (2006)

\bibitem{bf1} K. Bakke, L. R. Ribeiro, C. Furtado, Cent. Eur. J. Phys. {\bf{8(6)}} 893 (2010)

\bibitem{arf} G. B. Arfken, H. J. Weber, {\it{Mathematical Methods for Physicists}}, sixth edition (Elsevier Academic Press, New York, 2005)

\bibitem{pr} L. B. Castro, Eur. Phys. J. C {\bf{76}}, 61 (2016)

\bibitem{pr1} R. L. L. Vit\'oria, K. Bakke, Eur. Phys. J. C {\bf{78}} 175 (2018)

\bibitem{pr2} L. C. N. Santos, C. C. Barros Jr., Eur. Phys. J. C {\bf{78}}, 13 (2018)

\bibitem{pr3} K. Bakke, Ann. Phys. {\bf{346}}, 51 (2014)

\bibitem{me3} R. L. L. Vit\'oria, K. Bakke, Int. J. Mod. Phy. D {\bf{27}}, 1850005 (2018)

\bibitem{pr4} K. Bakke, Eur. Phys. J. B {\bf{85}}, 354 (2012)

\bibitem{pr5} K. Bakke, Int. J Theor. Phys. {\bf{54}}, 2119 (2015)

\bibitem{pr6} A. V. D. M. Maia, K. Bakke, Physica B {\bf{531}}, 213 (2018)

\bibitem{abr} M. Abramowitz, I. A. Stegum, {\it{Handbook of Mathematical Functions}} (Dover Publications Inc., New York, 1965)

\bibitem{kb} K. Bakke, Gen. Relativ. Gravit. {\bf{45}}, 1847 (2013)

\bibitem{sal} S. Salinas, {\it{Introdu\c c\~ao \`a F\'isica Estat\'istica}}, segunda edi\c c\~ao (Edusp, S\~ao Paulo, 2005)





\end{thebibliography}
\end{document}